\documentclass[12pt,a4paper,fleqn]{article}
\usepackage{amsmath,amssymb,amsfonts,amsthm,amsopn,graphicx}
\usepackage{mathrsfs} 
\usepackage{tikz} 
\usetikzlibrary{arrows}
\usepackage{hyperref}
\allowdisplaybreaks[3]
\numberwithin{equation}{section} 
\newtheorem{theorem}{Theorem}[section]

\newtheorem{stat}[theorem]{Statement}

\newcommand{\ds}{\displaystyle}
\def\EXP{\textrm{{\large e}}}
\newcommand{\ii}{{\sf i}}

\newcommand{\hl}{{^1\!\!/\!_2}}
\newcommand{\Om}{\textrm{\scalebox{1.2}{$\omega$}}}

\newcommand{\Si}{\textrm{\scalebox{1.1}{$\sigma$}}}

\def\uop{\boldsymbol{u}}
\def\wop{\boldsymbol{w}}
\def\vop{\boldsymbol{v}}
\def\Vop{\boldsymbol{V}}

\usepackage[body={15.5cm,22cm}]{geometry}
\begin{document}

\title{On Faddeev's Equation}
\author{Sergey Sergeev
\thanks{ 
Department of Fundamental and Theoretical Physics,
         Research School of Physics and Engineering,
    Australian National University, Canberra, ACT 0200, Australia}
    \thanks{
   Faculty of Science and Technology, 
   University of Canberra, Bruce ACT 2617, Australia
}}

\date{}

\maketitle

\begin{abstract}
Faddeev' equations are a set-theoretical and an operator forms of the star-triangle equation. 
Known solutions of the quantum star-triangle equation, related to the Faddeev equations, are 
based on various forms of the modular double of the Weyl algebra including its cyclic representation.
We show in this paper that Fadeev's equation also leads to a solution of the quantum star-triangle equation 
even in the case of a simple Weyl algebra with $|q|<1$.

This paper can be seen as an addendum to the recent paper \cite{BS}.
\end{abstract}



\section{Introduction}

There is a class of the solutions of the Baxter-Onsager star-triangle equation (a special form of the Yang-Baxter equation) where the two-spin Boltzmann weights depend on the difference of the ``spin variables''. This class includes the Chiral Potts model, the $\mathbb{Z}_N$ Zamolodchikov-Fateev model, Zamolodchikov's fishing net model, and a rather wide subclass of the models related to several forms of a modular double and a quantum dilogarithm \cite{Kash1}.
In the all above-mentioned cases the quantum star-triangle equation can be rewritten in a representation-independent form known as the Faddeev equation \cite{Faddeev}.

An essential ingredient of the Faddeev equation method is the Weyl algebra of observables. We show in this paper that the Faddeev method provides the solution to the star-triangle equation also in the case of an elementary representation of the simple Weyl algebra with generic $q$, $|q|<1$.

This paper is organised as follows. Firstly, in Section 2, we remind the reader the Faddeev equation method and the criteria of the existence of corresponding quantum star-triangle equation. Next, in Section 3, we construct a proper solution to Faddeev's recursion relations satisfying the criteria listed in Section 2. The resulting star-triangle equations in two dual forms are given in Section 4. The paper is concluded by some final remarks Section 5. In addition, the paper has a big Appendix containing a bare list of properties of a corresponding exactly solvable Ising-type model of the statistical mechanics.

\section{Faddeev equation's method}

Let $\uop$ and $\wop$ be the generating elements of a free associative algebra with unity. 
Let us define then two mappings on the space of functions of $\uop,\wop$: $\forall \; f\;=\;f(\uop,\wop)$,
\begin{equation}\label{map}
\left( \overline{\mathtt{r}}_{\lambda} \circ f\right) (\uop,\wop)\;=\;
f(\uop,\wop \frac{\lambda+\uop}{1+\lambda\uop})\;,\quad
\left( {\mathtt{r}}_{\lambda} \circ f\right) (\uop,\wop)\;=\;
f(\frac{1+\lambda\wop}{\lambda+\wop}\uop,\wop)\;,
\end{equation}
where $\lambda\in\mathbb{C}$. Then the set-theoretical Faddeev equation takes place:
\begin{equation}\label{fadeev}
(\overline{\mathtt{r}}_\lambda \circ ({\mathtt{r}}_{\lambda\mu} \circ ( \overline{\mathtt{r}}_\mu \circ f)))(\uop,\wop)\;=\;
({\mathtt{r}}_\mu \circ (\overline{\mathtt{r}}_{\lambda\mu} \circ ({\mathtt{r}}_\lambda \circ f)))(\uop,\wop)\;.
\end{equation}
Proof – the direct verification.

To bring the Faddeev equation to an operator form, one need to introduce the Weyl algebra,
\begin{equation}\label{weyl}
\uop\wop\;=\;q^2\wop \uop\;.
\end{equation}
Then one can re-define the functional map (\ref{map}) as the result of the adjoint action of corresponding operators,
\begin{equation}\label{conj}
(\overline{\mathtt{r}}_\lambda \circ f) \;=\; \overline{\mathtt{r}}_\lambda(\uop) \cdot f\cdot \overline{\mathtt{r}}_\lambda(\uop)^{-1}\;,\quad
({\mathtt{r}}_\lambda \circ f) \;=\; {\mathtt{r}}_\lambda(\wop) \cdot f \cdot {\mathtt{r}}_\lambda(\wop)^{-1}\;,
\end{equation}
so that the map (\ref{map}) is equivalent to the following difference relations,
\begin{equation}\label{req}
\frac{\overline{\mathtt{r}}_\lambda(q^2\uop)}{\overline{\mathtt{r}}_\lambda(\uop)}\;=\;
\frac{\lambda+\uop}{1+\lambda\uop}\;,\quad
\frac{{\mathtt{r}}_\lambda(q^2\wop)}{{\mathtt{r}}_\lambda(\wop)}\;=\;
\frac{\lambda+\wop}{1+\lambda\wop}\;.
\end{equation}
A solution of (\ref{req}) is a candidate for the operator Faddeev equation,
\begin{equation}\label{ybe}
\overline{\mathtt{r}}_\lambda(\uop)
\mathtt{r}_{\lambda\mu}(\wop) 
\overline{\mathtt{r}}_\mu(\uop)\;=\;
\mathtt{r}_\mu(\wop)
\overline{\mathtt{r}}_{\lambda\mu}(\uop)
\mathtt{r}_\lambda(\wop)\;.
\end{equation}
However, since the simple Weyl algebra has a big center (e.g. any periodical function $C(q^2\wop)=C(\wop)$ belongs to the center), equations (\ref{req}) do not guarantee the existence of (\ref{ybe}). Therefore some additional criteria could be taken into account.

Our conjecture is that the difference equations (\ref{req}), being considered together with the additional requirements
(inversion relations) 
\begin{equation}\label{crit}
\overline{\mathtt{r}}_\lambda(\uop) \overline{\mathtt{r}}_{1/\lambda}(\uop) \;=\; 1\;,\quad 
\mathtt{r}_{1/\lambda}(\wop) \mathtt{r}_\lambda(\wop)\;=\;1\;,
\end{equation}
are the sufficient conditions for the operator Faddeev equation (\ref{ybe}).

\section{Simple Weyl algebra and solution to (\ref{req},\ref{crit})}

Let us fix now (non-self-dual) representation of the simple Weyl algebra (\ref{weyl}):
\begin{equation}
\uop | n \rangle \;=\; | n \rangle q^{2n}\;,\quad
\wop | n \rangle \;=\; | n+1 \rangle\;.
\end{equation}
The left equation of (\ref{req}) can be solved straightforwardly,
\begin{equation}\label{ru1}
\overline{\mathtt{r}}_\lambda(q^{2n})=\overline{V}_\lambda(n),\;\;
\frac{\overline{V}_\lambda(n+1)}{\overline{V}_\lambda(n)} = \frac{\lambda+q^{2n}}{1+\lambda q^{2n}} \;\;
\Rightarrow\;\;
\overline{V}_\lambda(n) \;=\; \lambda^n \frac{(-q^{2n}\lambda;q^2)_\infty}{(-q^{2n}\lambda^{-1};q^2)_\infty}\;,
\end{equation}
so that the left inversion relation in (\ref{crit}) is satisfied in an elementary way. To solve the right equation in (\ref{req}), one can rewrite it as
\begin{equation}\label{weq}
(1+\lambda\wop)\uop\,\cdot\, \mathtt{r}_\lambda(\wop) \;=\;
\mathtt{r}_\lambda(\wop) \,\cdot\, (\lambda+\wop)\uop\;,
\end{equation}
so that the matrix element of (\ref{weq}) in between $\langle a |$ and $| b \rangle$, 
\begin{equation}
\langle a | \mathtt{r}_{\lambda}^{}(\wop) | b \rangle \;=\; \rho_\lambda^{-1} V_\lambda^{}(a-b)\;,
\end{equation}
satisfies
\begin{equation}\label{ru2}
V_\lambda(n) \;=\; -\, \frac{1-\lambda q^{2(n-1)}}{\lambda-q^{2n}}\, V_\lambda(n-1)\quad \Rightarrow\quad
V_\lambda(n) \;=\; (-\lambda)^{-n} \frac{(q^{2+2n}\lambda^{-1};q^2)_\infty}{(q^{2n}\lambda;q^2)_\infty}\;.
\end{equation}
The key observation made in \cite{BS} is that in order have the second inversion relation from (\ref{crit}) to be satisfied, one has to treat
\begin{equation}\label{halfint}
n\;\in\;\mathbb{Z}/2\quad \textrm{in the definition of}\quad V_\lambda(n),\overline{V}_\lambda(n)\;.
\end{equation}
This statement is based on the ``generalised Rogers-Ramanujan identity'' \cite{Kash1},
\begin{equation}\label{rw2v}
\mathtt{r}_\lambda(\vop)\;\stackrel{def}{=}\; \frac{1}{\Si(\lambda)\theta_4}\;
 \sum_{n\in\mathbb{Z}} \vop^{n} 
\underbrace{ 
 (\ii\lambda^{-^1\!\!/\!_2})^n 
\frac{(q^{2+n}\lambda^{-1};q^2)_\infty}{(q^n\lambda;q^2)_\infty}
}_{V_\lambda(^n\!/\!_2)}\;=\;
\Si(\ii\lambda^{-^1\!\!/\!_2}\vop) 
\Si(\ii q \lambda^{-^1\!\!/\!_2}\vop^{-1})
\;,
\end{equation}
which follows from the standard Rogers-Ramanujan identity for $_1\psi_1$ straightforwardly, where 
\begin{equation}\label{Si}
\Si(x)\;\stackrel{def}{=}\;\frac{(-q/x;q)_\infty}{(x;q)_\infty}\;,\quad 
\theta_4\;=\;\frac{(q;q)_\infty}{(-q;q)_\infty}\;,\quad \varrho_\lambda\;=\;\theta_4\Si(\lambda)\;,
\end{equation}
and $\vop$ is a ``root of $\wop$'':
\begin{equation}\label{v}
\wop\;=\;\vop^2\;,\quad 
\vop | a \rangle \;=\; | a+{^1\!\!/\!_2} \rangle\;.
\end{equation}
The inversion relation $\mathtt{r}_\lambda(\vop)\mathtt{r}_{^1\!/\!_\lambda}(\vop)\;=\;1$ follows from the property $\Si(x)\Si(-q/x)=1$. Numerical tests allow one to come to
\begin{stat}
Equation (\ref{ybe}) with the definitions (\ref{ru1},\ref{rw2v})
is satisfied in matrix elements (\ref{halfint}).
\end{stat}

\section{Star-triangle equations}

The Faddeev equation (\ref{ybe}) is the operator form of the Baxter star-triangle equation. Namely, the matrix element of (\ref{ybe}) is equivalent (after a minor change) to 
\begin{equation}\label{str}
\begin{array}{l}
\ds \theta_4\, \frac{\Si(\lambda)\Si(\mu)}{\Si(\lambda\mu)}  \;\overline{V}_\lambda(a-c) V_{\lambda\mu}(a-b) \overline{V}_\mu(b-c)\;=\\
\\
\ds \qquad\qquad\qquad =\;
\sum_{d\in\mathbb{Z}}
V_\mu(a-d) \overline{V}_{\lambda\mu}(d-c) V_{\lambda}(d-b)\;,
\end{array}
\end{equation}
where the weight are slightly re-defined,
\begin{equation}\label{VV}
\overline{V}_\lambda(a)\;\stackrel{def}{=}\;\lambda^{a/2} \frac{(-q^{1+a}\lambda;q^2)_\infty}{(-q^{1+a}\lambda^{-1};q^2)_\infty}\;,\quad
V_\lambda(a)\;\stackrel{def}{=}\;\left(-\frac{q}{\lambda}\right)^{a/2} \frac{(q^{2+a}\lambda^{-1};q^2)_\infty}{(q^a\lambda;q^2)_\infty}\;,
\end{equation}
and extra multipliers are defined in by (\ref{Si}). The advantage of the new definition (\ref{VV}) is the symmetry,
\begin{equation}
V_\lambda(-n) \;=\; V_\lambda(n)\;,\quad 
\overline{V}_\lambda(n)\;=\;V_{-q/\lambda}(n)\;.
\end{equation}
The Fourier transform allows one to rewrite the star-triangle equation in its dual form of another star-triangle equation. Let us define 
\begin{equation}\label{Om}
\Om_\lambda(\varphi)\;\stackrel{def}{=}\;\frac{\Si(\lambda)}{\theta_4} \sum_{a\in\mathbb{Z}} \overline{V}_\lambda(a) \EXP^{\ii a \varphi}\;=\;
\Si(\lambda^{^1\!\!/\!_2}\EXP^{\ii\varphi})\Si(\lambda^{^1\!\!/\!_2}\EXP^{-\ii\varphi})\;,
\end{equation}
where the result of summation corresponds to the application of the ``generalised Rogers-Ramanujan identity''
(\ref{rw2v}), and
\begin{equation}
\overline{\Om}_\lambda(\varphi)\;\stackrel{def}{=}\;\Om_{-q/\lambda}(\varphi)\;.
\end{equation}
Using these notations and making the Fourier transform of (\ref{str}), one ontains
\begin{equation}\label{STR3}
\begin{array}{l}
\ds \frac{\Si(\lambda)\Si(\mu)}{\Si(\lambda\mu)}\;
\overline{\Om}_\mu(\alpha) \, \Om_{\lambda\mu}(\alpha-\beta) \, \overline{\Om}_\lambda(\beta)\;=\\
\\
\ds \qquad \qquad\qquad =\; \frac{\theta_4}{2\pi} \int_{-\pi}^{\pi} \, d\varphi \,
\Om_\lambda(\alpha-\varphi)\;
\overline{\Om}_{\lambda\mu}(\varphi)\;
\Om_\mu(\varphi-\beta)\;.
\end{array}
\end{equation}
This star-triangle equation can be rewritten in Spiridonov's style,
\begin{equation}\label{tozh}
\frac{\theta_4}{2\pi\ii} \oint \; \frac{dv}{v} \;
\prod_{j=1}^3 \frac{\Si(a_jv)}{\Si(b_jv)}\;=\; 
\left( \prod_{i,k=1}^3 \Si(\frac{b_j}{a_k}) \right)^{-1}\;,\quad
\frac{b_1b_2b_3}{a_1a_2a_3}\;=\;q^2\;,
\end{equation}
where the contour must be chosen so that 
\begin{equation}
\max\left( |q/b_j| \right)\,<\,|v|\,<\,\min \left( 1/|a_j| \right)\;.
\end{equation}

\section{Conclusion}

The solution of the star-triangle equation obtained in this paper has the physical regime for statistical mechanics (when all weights are positive),
\begin{equation}
0<-q<\lambda<1\;.
\end{equation}
Starting from the star-triangle equation, one can construct in usual way transfer matrices ($Q$-operators), derive Baxter's $TQ$ equation \cite{Baxterbook} and other ingredients of an integrable model \cite{BS}. The reader can find brief results in the Appendix. One can calculate partition function, local state probability, and construct the spectrum of the model by means of the Sklyanin method of the functional Bethe Ansatz \cite{SS}. The model in the critical points becomes the Zamolodchikov $D=1$ fishing net model \cite{Fish} in the limit $-q\to 1$, or the $\mathbb{Z}_N$ Fateeev-Zamolodchikov model \cite{FZ} in the limit $q^N\to 1$.

Concluding this paper, we'd like to note that expressions like (\ref{VV}) appeared in Woronovich' studies \cite{Wor}, they appeared in Dimofte-Gaiotto-Gukov's  paper \cite{DGG} under the name ``index for an ideal tetrahedron for $3d$ Super Conformal Field Theory''. 
These expressions were systemised by Stavros Garoufalidis and Rinat Kashaev in \cite{Kash1,Kash2} in application to the theory of quantum dilogarithms and pentagon equations. In particular, Kashaev and Garoufalidis used extensively the function $\Si$, they derived the ``generalised Rogers-Ramanujan identity'' (\ref{rw2v}) and they derived the identity (\ref{tozh}) under the name ``beta Pentagon identity''.

\vspace{1cm}
\noindent
\textbf{Acknowledgements.} The author would like to thank R. Kashaev, V. Bazhanov and V. Mangazeev for valuable discussions. The work was partially supported by the ARC grant DP190103144.

\appendix

\section{A digest of the integrability}

We collect here some general results for the obtained model of statistical mechanics.

\subsection{Initial definitions}

Physical regime is
\begin{equation}
0<-q<x<1\;.
\end{equation}
The discrete spin weight are given by
\begin{equation}\label{weights-V}
V_x(n)\;=\;\left(-\frac{q}{x}\right)^{n/2} \frac{(q^{2+n}x^{-1};q^2)_\infty}{(q^nx;q^2)_\infty}\;,\quad
\overline{V}_x(n)\;=\;x^{n/2} \frac{(-q^{1+n}x;q^2)_\infty}{(-q^{1+n}x^{-1};q^2)_\infty}
\end{equation}
Their symmetries are given by
\begin{equation}
V_x(-n)\;=\;V_x(n)\;,\quad \overline{V}_x(n)\;=\;V_{-q/x}(n)\;.
\end{equation}
An important function is defined by 
\begin{equation}
\Si(v)\;=\;\frac{(-q/v;q)_\infty}{(v;q)_\infty}\;,\quad \Si(v) \Si(-q/v)\;=\;1\;.
\end{equation}
Normalization:
\begin{equation}
\theta_4\;=\;\frac{(q;q)_\infty}{(-q;q)_\infty}\;=\;\sum_{n\in\mathbb{Z}} (-q)^{n^2}\;.
\end{equation}
The Boltzmann weight in the compact representation are defined by
\begin{equation}\label{weights-O}
\Om_x(\varphi)\;=\;\Si(x^{\hl}\EXP^{\ii\varphi}) \Si(x^{\hl}\EXP^{-\ii\varphi})\;,\quad
\overline{\Om}_x(\varphi)\;=\;\Om_{-q/x}(\varphi)\;.
\end{equation}
The Fourier transforms for the weights are given by
\begin{equation}\label{Furier}
\left\{
\begin{array}{l}
\ds \Om_x(\varphi)\;=\;\frac{\Si(x)}{\theta_4}\,  
\sum_{n\in\mathbb{Z}} \, \overline{V}_x(n) \, \EXP^{\ii\varphi n}\;,\\
\\
\ds 
V_x(n) \;=\; \theta_4 \Si(x) \ds \int_{-\pi}^{\pi} \frac{d\varphi}{2\pi}\;
\overline{\Om}_x(\varphi) \, \EXP^{-\ii \varphi n}
\end{array}\right.
\end{equation}
The inversion relations are given by
\begin{equation}\label{inv}
\left\{
\begin{array}{l}
\ds \sum_{n\in\mathbb{Z}}\; V_x(a-n)\, V_{1/x}(n-b)\;=\; \theta_4^2 \Si(x)\Si(1/x) \delta_{a,b}\;,\\
\\
\ds V_x(n)\,V_{q^2/x}(n)\;=\;1\;,\\
\\
\ds 
\theta_4^2\int_{-\pi}^{\pi} \frac{d\varphi}{2\pi} \; \Om_x(\alpha-\varphi) \Om_{1/x}(\varphi-\beta) \;=\; \Si(x)\Si(1/x)\,  2\pi 
 \, \delta(\alpha-\beta)\;,\\
\\
\Om_x(\varphi) \, \Om_{q^2/x}(\varphi)\;=\;1\;.
\end{array}\right.
\end{equation}
Some limits,
\begin{equation}
\lim_{x\to 1} \; \frac{1}{\theta_4\Si(x)} V_x(n)\;=\;\delta_{n,0}\;,\quad
\lim_{x\to 1} \; \theta_4 \frac{\Om_x(\alpha)}{2\pi \Si(x)}\;=\;\delta(\alpha)\;.
\end{equation}
The star-triangle relations:
\begin{equation}
\left\{
\begin{array}{l}
\ds \theta_4 \frac{\Si(x)\Si(y)}{\Si( xy )}\;
\overline{V}_x(a) \, V_{xy}(a-b) \, \overline{V}_y(b) \;=\\
\\
\ds \qquad\qquad 
=\; \sum_{n\in\mathbb{Z}}\; V_y(a-n) \, \overline{V}_{xy}(n) \, V_x(n-b)\;,\\
\\
\\
\ds \frac{\Si(x)\Si(y)}{\Si( xy )}\;
\overline{\Om}_x(\alpha) \, \Om_{xy}(\alpha-\beta) \, \overline{\Om}_y(\beta) \;=\\
\\
\ds \qquad\qquad 
=\; \theta_4 \int_{-\pi}^{\pi} \frac{d\varphi}{2\pi}\; 
\Om_y(\alpha-\varphi) \, \overline{\Om}_{xy}(\varphi) \, \Om_x(\varphi-\beta)\;.
\end{array}
\right.
\end{equation}

\subsection{The Baxter equation}

A ``simplified'' $\hat{Q}$-operator can be defined by
\begin{equation}\label{Q}
\langle a | \hat{Q}(\lambda;x,x') | b \rangle\;=\;
\prod_{n=1}^N \; \overline{V}_{x/\lambda}(a_n-b_n) 
\overline{V}_{\lambda/x'}(a_{n+1}-b_n) V_{x/x'}(a_n-a_{n+1})\;.
\end{equation}
This is in fact the transfer matrix for a triangular lattice. Such simple form of $\hat{Q}$ operator is valid for a homogeneous chain only. This $\hat{Q}$ operator is normalised by
\begin{equation}
\lim_{\lambda\to -x/q} \left( \frac{\Si(x/\lambda)}{\theta_4}\right)^N \langle a | \hat{Q}(\lambda;x,x') | b \rangle \;=\;
\prod_{n=1}^N \; \delta_{a_n,b_n}
\end{equation}
The other limit, $\lambda\to-qx'$, gives a shift operator. This $\hat{Q}$ - operator is related to an auxiliary transfer matrix constructed with the help of the following $L$ -- operator,
\begin{equation}\label{Lop}
\hat{L}(\lambda;x,x')\;=\;
\left(
\begin{array}{cc}
\ds [\frac{(xx')^{\hl}}{\lambda} \vop] & \ds [ \left(\frac{x'}{x}\right)^{\hl}\vop^{-1}]\uop \\
\\
\ds [ \left(\frac{x'}{x}\right)^{\hl}\vop] \uop^{-1} & \ds [\frac{(xx')^{\hl}}{\lambda}\vop^{-1}]
\end{array}
\right)
\end{equation}
where the simple Weyl algebra is considered in the following representation:
\begin{equation}\label{Weyl1}
\uop | a \rangle \;=\; | a \rangle \; q^{a}\;,\quad 
\vop | a \rangle \;=\; | a+1 \rangle\;,\quad \langle a | b \rangle \;=\; \delta_{a,b}\;,
\end{equation}
and
\begin{equation}
[x]\;=\;x-x^{-1}\;.
\end{equation}
The auxiliary tranfer-matrix is defined by
\begin{equation}
\hat{T}(\lambda)\;=\;\textrm{Trace}\; 
\hat{L}_1(\lambda;x,x') \hat{L}_2(\lambda;x,x') \cdots \hat{L}_N(\lambda;x,x')
\end{equation}
All transfer matrices commute,
\begin{equation}
[\hat{T}(\lambda),\hat{T}(\lambda')]\;=\;
[\hat{T}(\lambda),\hat{Q}(\lambda')]\;=\;
[\hat{Q}(\lambda),\hat{Q}(\lambda')]\;=\;0\;,
\end{equation}
and the Baxter equation has the following form:
\begin{equation}\label{tq}
\hat{T}(\lambda)\hat{Q}(\lambda)\;=\; [\frac{x'}{\lambda}]^N \hat{Q}(q\lambda) + [\frac{x}{\lambda}]^N \hat{Q}(q^{-1}\lambda)\;.
\end{equation}
In addition, there is the Wronskian relation,
\begin{equation}\label{wrq}
\hat{Q}(q\lambda)\hat{Q}(-\lambda) \;-\;
\hat{Q}(\lambda)\hat{Q}(-q\lambda)\;=\;
\left(\theta_4^2\frac{\Si(-x'/\lambda)}{\Si(-x/\lambda)}\right)^N\;-\;
\left(\theta_4^2\frac{\Si(x'/\lambda)}{\Si(x/\lambda)}\right)^N\;.
\end{equation}

\subsection{Fourier transform}

Define the following kernel for the Fourier transform:
\begin{equation}\label{Fk}
\langle b |\mathscr{F} | \beta \rangle \;=\; \exp\left\{ -\ii \sum_{n=1}^N b_n (\beta_n - \beta_{n+1}) \right\}\;.
\end{equation}
Then the following identity holds:
\begin{equation}\label{QQ}
\sum_{b}\;  \langle a | \hat{Q}(\lambda;x,x') | b \rangle \, \langle b | \mathscr{F} | \beta \rangle \;=\; 
\theta_4^{2N} \; \int \mathscr{D}\alpha \; \langle a | \mathscr{F} | \alpha \rangle \;\langle \alpha | \check{Q}(\lambda;x,x') | \beta \rangle
\end{equation}
where
\begin{equation}
\mathscr{D}\alpha \;=\; \prod_{n=1}^N \frac{d\alpha_n}{2\pi}\;,
\end{equation}
and the kernel of the left $\hat{Q}$-operator is given by (\ref{Q}), while the kernel of the right $\check{Q}$-operator is given by 
\begin{equation}\label{Q2}
\langle \alpha | \check{Q}(\lambda;x,x') |\beta \rangle \;=\;
\prod_{n=1}^N\; \overline{\Om}_{x/\lambda}(\alpha_n-\beta_n) 
\overline{\Om}_{\lambda/x'}(\alpha_{n+1}-\beta_n) \Om_{x/x'}(\beta_n-\beta_{n+1})
\end{equation}
It is important to note, operator (\ref{Fk}) is non-local. Also, operator (\ref{Fk}) is a projection operator, 
it is invariant with respect to a simultaneous shift $\beta_n\to\beta_n+1$ or $b_n\to b_n+1$. As the result, operator $\check{Q}$ 
is related to 
\begin{equation}
\check{L}(\lambda;x,x')\;=\;
\left(
\begin{array}{cc}
\ds [\frac{(xx')^{\hl}}{\lambda}\vop] & \ds \uop [\left(\frac{x'}{x}\right)^{\hl}\vop] \\
\\
\ds \uop^{-1} [\left(\frac{x'}{x}\right)^{\hl}\vop^{-1}] & \ds [\frac{(xx')^{\hl}}{\lambda}\vop^{-1}]
\end{array}
\right)\;,
\end{equation}
where the representation of the Weyl algebra is different,
\begin{equation}\label{Weyl2}
\uop | \beta \rangle \;=\; | \beta \rangle \EXP^{2\ii\beta}\;,\quad
\vop | \beta \rangle \;=\; | \beta+\eta\rangle\;,\quad q\;=\;\EXP^{2\ii\eta}\;.
\end{equation}
The analogous relation for the auxiliary transfer matrices is 
\begin{equation}\label{TT}
\sum_b \langle a | \hat{T}(\lambda;x,x') | b \rangle \; 
\langle b | \mathscr{F}  | \beta \rangle \;=\; \int \mathscr{D}\alpha \langle a | \mathscr{F} | \alpha \rangle \;
\langle \alpha | \check{T}(\lambda;x,x') | \beta \rangle\;,
\end{equation}
so that the Baxter equation for $\check{T}$, $\check{Q}$ coincides with the Baxter equation (\ref{tq}). 
However, the difference between $\hat{T}(\lambda)$ and $\check{T}(\lambda)$ is to be mentioned. In both cases 
\begin{equation}
T(\lambda) \;=\; \left(\frac{(xx')^{\hl}}{\lambda}\right)^N (\Vop+\Vop^{-1}) + \cdots +
\left(- \frac{\lambda}{(xx')^{\hl}}\right)^N (\Vop+\Vop^{-1})\;,
\end{equation}
where $\ds \Vop\;=\;\vop_1\vop_2\cdots\vop_N$ is an analogue of the total spin operator for the six-vertex model. 
The representation (\ref{Weyl1}) provides $\Vop\;=\;\EXP^{\ii\varphi_0}$, while the representation (\ref{Weyl2}) provides $\Vop=q^{m_0/2}$. The relations (\ref{QQ}) and (\ref{TT}) establish the equivalence of the spectra of $\hat{T}$ and $\check{Q}$ and the spectra of $\check{T}$ and $\check{Q}$ in the sector $\varphi_0=m_0=0$ only.

Note that the simple Fourier transform of (\ref{Q}), which corresponds to the simple basis transformation,
\begin{equation}
\langle \varphi | a \rangle \;=\; \EXP^{\ii\varphi a}\;,\quad \langle \varphi | \psi \rangle \;=\; \sum_a \EXP^{\ii\varphi a}\;\langle a | \psi \rangle \;,
\end{equation}
gives
\begin{equation}
\begin{array}{l}
\ds \sum_{a,b} \EXP^{\ii \alpha a - \ii \beta b} \;
\langle a | \hat{Q}(\lambda;x,x') | b \rangle \;=\\
\\
\ds =\; 
\theta_4^{2N} \int \prod_n d\varphi_n \delta(\varphi_{n+1}-\varphi_n-\beta_n+\alpha_{n+1})
\overline{\Om}_{x/\lambda}(\beta_n+\varphi_n) \overline{\Om}_{\lambda/x'}(\varphi_n)
\Om_{x/x'}(\beta_n)
\end{array}
\end{equation}
where the left hand side contains one additional integral with respect to (\ref{Q2}).

\subsection{Partition functions}

Partition functions per one site are given by
\begin{equation}
\sum_n\;\to\; \theta_4\;,\quad
\int \frac{d\varphi}{2\pi}\;\to\; \frac{1}{\theta_4}\;.
\end{equation}
Partition function per one two-spin weight $V_x(n)$ or $\Om_x(\varphi)$ satisfy the following functional equations,
\begin{equation}
z_x z_{1/x} \;=\; \Si(x) \Si(1/x)\;,\quad z_x z_{q^2/x} \;=\;1\;,\quad
z_x \;=\; \Si(x) z_{-q/x}\;.
\end{equation}
Partition function per one weight is given then
\begin{equation}
\log z_x\;=\;\sum_{n=1}^\infty \frac{x^{n} - (q^2/x)^{n}}{n(1-q^n)(1+(-q)^n)}\;.
\end{equation}
In the discrete spin formulation, the probability to have spin $m$ in the middle of the lattice with zero spins on infinity is given by
\begin{equation}\label{LSP}
P(m)\;=\;\int_{-\pi}^{\pi}\; \frac{d\varphi}{2\pi} \; \EXP^{\ii\varphi m}\;
\frac{(-q;q^2)_\infty^2}{(-q\EXP^{\ii\varphi},-q\EXP^{-\ii\varphi};q^2)_\infty}
\end{equation}
Formula (\ref{LSP}) can be obtained by the Corner Transfer Matrix method \cite{Baxterbook}.

\end{document}